\documentclass[
	aps,reprint,floats,floatfix,amsmath,superscriptaddress,longbibliography
]{revtex4-2}
\usepackage[utf8]{inputenc}

\usepackage{graphicx}
\usepackage[per-mode=symbol,inter-unit-product=\cdot]{siunitx}
\usepackage{hyperref}
\usepackage{amsmath}

\date{\today}

\begin{document}

\title{$N$ Scaling of Large-Sample Collective Decay in Inhomogeneous Ensembles}

\author{Sergiy Stryzhenko}
\affiliation{Institut f\"{u}r Angewandte Physik, Technische Universit\"{a}t Darmstadt, Hochschulstra\ss e 6, 64289 Darmstadt, Germany}
\affiliation{Institute of Physics, National Academy of Science of Ukraine, Nauky Avenue 46, Kyiv 03028, Ukraine}

\author{Alexander Bruns}
\affiliation{Institut f\"{u}r Angewandte Physik, Technische Universit\"{a}t Darmstadt, Hochschulstra\ss e 6, 64289 Darmstadt, Germany}

\author{Thorsten Peters}
\affiliation{Institut f\"{u}r Angewandte Physik, Technische Universit\"{a}t Darmstadt, Hochschulstra\ss e 6, 64289 Darmstadt, Germany}
\email[]{thorsten.peters@physik.tu-darmstadt.de}

\date{\today}

\begin{abstract}
We experimentally study collective decay of an extended disordered ensemble of $N$ atoms inside a hollow-core fiber. We observe up to $300$-fold enhanced decay rates, strong optical bursts and a coherent ringing. Due to inhomogeneities limiting the synchronization of atoms, the data does not show the typical scaling with $N$. We show that an effective number of collective emitters can be determined to recover the $N$ scaling known to homogeneous ensembles over a large parameter range. This provides physical insight into the limits of collective decay and allows for its optimization in extended ensembles as used, e.g., in quantum optics, precision time-keeping or waveguide QED.
\end{abstract}

\maketitle

\section{Introduction}
The rate $\Gamma$ at which an individual two-level system (TLS) emits
radiation depends on its coupling to the available vacuum modes and can thus
be altered when, e.g., placed inside a cavity \cite{HK89}.
For an ensemble of $N$ \textit{identical} TLSs that are
dipole-dipole coupled to each other via the emitted
radiation, the decay rate can change significantly even
without a cavity, as first noted by Dicke \cite{D54}.
Here, coherence can potentially build up between the emitters,
leading to their spontaneous synchronization \cite{BGP17}, and an enhanced collective decay rate $\Gamma_N>\Gamma$.
This is termed superradiance (SR) \cite{GH82,GRK16}, if there is some initial
coherence present, and superfluorescence (SF) \cite{BL75} if the ensemble is
initially completely inverted and coherence is initiated by vacuum
fluctuations.
Since its first observation \cite{SHM73}, studies of enhanced collective decay
were reported for various experimental platforms such as free-space
\cite{OMM14,BZB16,RKH16,AKK16} and waveguide-coupled atomic ensembles
\cite{GHH15,SBF17,OYV19,PBJ22}, solid-state materials \cite{CZW16}, superconducting
qubits \cite{MAE14} and biological systems \cite{MAV97}.
Besides being interesting due to its fundamental nature, applications of collective decay can be found, e.g., in quantum optics \cite{AMA17}, waveguide QED \cite{SPI23}, astrophysics \cite{HMR18}, atomic clocks
\cite{MYC09,NWC16}, and chemistry \cite{MAV97,FRN18}.

SR/SF can be observed in small, dense samples \cite{FGR21} as well as in large, dilute samples, where propagation effects have to be considered. 
Both cases exhibit several $N$ scaling signatures \cite{GH82}. 
In the small-sample case, when $N$ emitters are confined to a volume of dimensions smaller than defined by the emitted wavelength $\lambda$, the initial decay rate is enhanced and given by $\Gamma_N = N \Gamma$.
In the large-sample case, where the $N$ emitters are distributed in a volume $>\lambda^3$, excitation of a homogeneously delocalized coherence is crucial, taking into account a position-dependent phase factor. 
In the regime of linear-optics SR this is reflected in the timed-Dicke state \cite{SFO06}.
Due to the extended ensemble volume, a geometric factor $\mu$ has to be introduced, describing the fraction of spontaneous radiation that can coherently couple to the ensemble of $N$ emitters in the detected mode \cite{GH82}.
The initial decay rate is then $\Gamma_N = \mu N \Gamma$, which is a factor $\mu$ smaller than in the small-sample case. $\mu$ and $\mu N$ ($\equiv N_\mu$) are the single-atom and collective cooperativities \cite{TLS11}, respectively, and $N_\mu$ becomes the new scaling parameter (see, e.g., \cite{PBK08,FGF23}).
Besides enabling collectively-enhanced decay rates, also bursts can be generated in SR/SF if the ensemble is initially inverted, with a delay $t_D$ of the burst after the inversion.
The photon rate of the burst evolves nonlinearly, reaching its peak at $t_D$ with a value $\propto N^2$. This burst occurs when $t_D, \Gamma_N^{-1} \ll \gamma^{-1}$, where $\gamma$ is the decoherence rate of the TLSs, which defines a threshold condition.
We note that recently a general condition for the occurrence of bursts in waveguide QED systems was derived \cite{CMZ23}.

As homogeneous excitation conditions in large samples can be challenging to realize (see e.g. \cite{CZW16,BJB17}), the effect of inhomogeneities on collective scattering are still investigated (see, e.g.,
\cite{JRL14,IMA16,AP21})--specifically in optically-dense ensembles \cite{BBP13,PBK08,BBL10,BPC11,AKK16,WCK21}.
In such inhomogeneous conditions, collective decay occurs only in sub-systems of the ensemble for which
the threshold condition is surpassed, while each of them radiates independently \cite{IMA16}.
This leads to a breakdown of the textbook $N$ scaling \cite{KH98,PBK08,AKS10,AKK16,ASA18}, potentially hiding the signature of collective decay.
As noted by Arecchi and Courtens \cite{AC70}, however, a maximum cooperation number (MCN) $N_\text{mc} \leq N_\mu$ can be determined when propagation losses, delays and inhomogeneous broadening become relevant, which replaces $N_\mu$ in the equations characterizing collective emission.
This concept of a MCN allows for understanding the physical limits to collective decay and has been applied, e.g., to cases where propagation time through the ensemble is longer than the timescale of cooperation build-up \cite{KH98,AKS10}.
As we will show in the following, a MCN can also be determined in cases where spectral and spatial inhomogeneities are relevant, thereby recovering the scaling expected for homogeneous ensembles.

In this paper we experimentally study large-sample SF for a broad range of inhomogeneous excitation conditions.
For $N \leq 2.2 \times 10^5$ atoms confined within a hollow-core photonic bandgap fiber (HCPBGF), we determine collective cooperativities up to $\sim 300$.
Due to intrinsic spatial and spectral inhomogeneities, the experimental data does not exhibit a textbook $N$ scaling.
However, by accounting for inhomogeneities in a simple model we determine a MCN which recovers the textbook scaling over a large range of experimental parameters when dispersion is negligible.
Our model allows for physical insight into the limits of collective decay in large ensembles, which is relevant to understand and optimize collective effects in, e.g., free-space \cite{OMM14,BZB16,RKH16,AKK16,GAK16} or waveguide-coupled ensembles \cite{SBF17,NSY18,DAV19,OYV19,LTB22,SPI23}, or solids \cite{CZW16}.

The paper is organized as follows: In Sec.~\ref{sec:model} we describe the model system used to study SF. Section~\ref{sec:expsetup} presents an overview of the experimental setup followed by the experimental results in Sec.~\ref{sec:expstudies}. In Sec.~\ref{sec:mcn} we then describe the model used to determine a MCN and compare its results to the experimental data. Following a discussion of the MCN model in Sec.~\ref{sec:discussion}, we conclude with a summary and outlook in Sec.~\ref{sec:summary}.

\section{Model System}\label{sec:model}
\begin{figure}[tb]
\includegraphics[width=0.8\columnwidth]{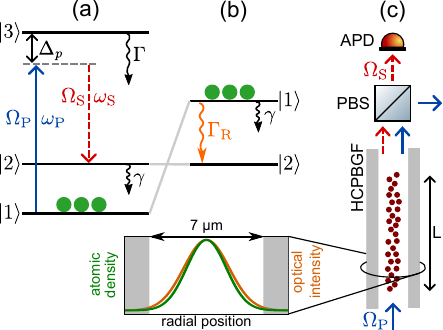}
\caption{
	\label{fig:levelscheme}
	(a)~Simplified three level system with
	$|1\rangle \,\hat{=}\,  {}^2S_{1/2} \, F{=}1$,
	$|2\rangle \,\hat{=}\, {}^2S_{1/2} \, F{=}2$, and
	$|3\rangle \,\hat{=}\, {}^2P_{1/2} \, F'{=}2$.
	The excited state decay rate is $\Gamma=2\pi \cdot \qty{5.75}{\MHz}$ \cite{Steck2019}.
	(b)~Corresponding effective TLS comprised of the ground states
	$|1\rangle$, $|2\rangle$.
	(c)~Simplified experimental setup.}
\end{figure}

We study SF for a disordered, cold ensemble of $^{87}$Rb atoms loaded into a HCPBGF
(NKT Photonics, HC-800-02) \cite{BHP14,PYH21}. 
The ensemble can be considered dilute as even at the peak atomic density $\mathcal{N}\sim 10^{12}~$cm$^{-3}$ we have $\lambda^3 \mathcal{N}<1$.
The single-mode waveguide provides long-range coupling between distant atoms, thereby facilitating the observation of collective effects \cite{SBF17,OMG19} while avoiding the use of a multi-mode theory to analyze the data \cite{OYV19}.
The atomic ensemble possesses an inhomogeneous radial density profile due to the guiding potential, which is of similar width as the excitation field [see Fig.~\ref{fig:levelscheme}(c)] \cite{PYH21}.
Moreover, as the ensemble exhibits a large optical depth (OD) \cite{BHP14}, attenuation of the excitation field has to be considered.
Thus, the system shows large radial and longitudinal inhomogeneities.
Radiation trapping is expected to be negligible due to an ensemble aspect ratio of order $10^4:1$. 
In contrast to typical realizations of SR/SF, where the excited state has a finite lifetime of $\Gamma^{-1}$, we here study the decay of an effective TLS comprised of two
long-lived ground states [see Fig.~\ref{fig:levelscheme}(a,b)]
\cite{RM81,WY05}. 
An off-resonant excitation field (termed pump) scatters off the ensemble prepared in $|1\rangle$.
Once the pump is on, spontaneous Raman scattering induced by vacuum fluctuations creates a delocalized coherence between states $|1\rangle$ and $|2\rangle$. This initiates the decay, followed by a buildup of radiation bursts in the Stokes field.
We note that this quasi-continuous, off-resonant driving of the system is different from the strong, resonant driving in \cite{FGF23}.
The single-atom scattering rate of such an effective TLS is for large detunings given by
\begin{equation}
    \Gamma_R = R_B \Gamma \frac{\Omega_p^2}{4\Delta^2},
    \label{eqn:scattering_rate}
\end{equation}
where $R_B(=0.5)$ is the branching ratio into the ground state $|2\rangle$,
$\Gamma$ is the excited state decay rate, and $\Omega_p$ is the Rabi frequency
of the pump field effectively detuned from the transition
$|1\rangle\leftrightarrow |3\rangle$ by $\Delta=\Delta_p+2S$.
Here, $\Delta_p$ is the externally applied detuning and
$2S=\Omega_p^2/(2\Delta_p)$ is the AC Stark shift of the $|1\rangle \leftrightarrow |3\rangle$ transition frequency
due to the strong and detuned pump field.

Studying SF in an effective TLS has several advantages over real TLSs.
First, $\Gamma_R$ can be tuned by the intensity and detuning $\Delta_p$ of the pump field.
Second, $\Gamma_R$ can be made much smaller than the excited state decay rate
$\Gamma$, which sets the width of the dispersive resonance.
Thus, even for large decay rates $\Gamma_N \gg \Gamma_R$, we still can have
$\Gamma_N < \Gamma$, and dispersive effects as discussed in \cite{WCK21} are more likely to be negligible.
This leads to a significant simplification when modeling the MCN.
Third, the modulation bandwidth of the excitation field has to be larger than
$\Gamma_N$.
This can be technically challenging if the single-atom decay rate $\Gamma$ is already quite large \cite{PBK08}.

\section{Experimental Setup}\label{sec:expsetup}
Figure~\ref{fig:levelscheme}(c) shows our simplified experimental setup.
A detailed description including the procedure to load atoms into the HCPBGF can be found in \cite{PYH21}.
Therefore, we here only give a brief summary.
First, we load a magneto-optical trap with around $10^7$ $^{87}$Rb atoms.
Then, we shift the atom cloud towards the HCPBGF tip while compressing it
radially using magnetic fields.
Up to $2.5\%$ of the atoms are then loaded into a red-detuned Gaussian-shaped
far-off-resonant trap (FORT) emerging from the HCPBGF with a numerical aperture $\textrm{NA} = 0.092(6)$.
The FORT guides the atoms radially inside the fiber, while they are basically
free-falling in the longitudinal (vertical) direction, resulting in a radial $1/e$ half-width $\sigma_a \sim \qty{1.7}{\um}$ and a length $L\sim\qty{3}{\cm}$ of the ensemble.
We prepare the atoms in state $|1\rangle$ by optical pumping without specific population of Zeeman levels.
The linear-polarized pump beam is launched into the fiber in such a way that its polarization is maintained at the exit to allow for efficient polarization filtering \cite{PWN20}.
Its temporal intensity is controlled by an acousto-optic modulator with a rise time of $\tau_p=\qty{130(5)}{\ns}$.
The radial pump intensity has a near-Gaussian profile with a $1/e^2$ half-width of $\sigma_p\sim \qty{2.75(20)}{\um}$. The shot-to-shot intensity fluctuations are $\lesssim 5\%$.
We detect the transmitted light using an avalanche photodiode (APD) and a digital oscilloscope.

\section{Experimental Studies}\label{sec:expstudies}
\begin{figure}[tb]
    \centering
    \includegraphics[width=\columnwidth]{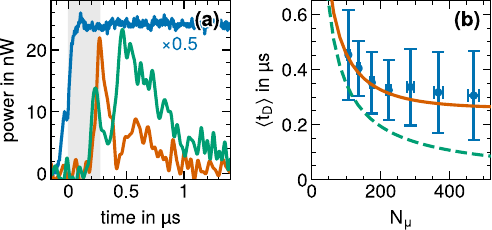}
    \caption{
    	(a)~Exemplary traces of the transmitted pump power without
    	atoms (blue, 16 averages, scaled down by factor $0.5$) and the Stokes power with
    	atoms loaded into the HCPBGF (single-shot data).
    	The peak pump Rabi frequency is $\Omega_p^{(0)}=6.4\Gamma$,
    	$\Delta_p=18.4\Gamma$ with $N=\num{83e3}$ (orange) and
    	$N=\num{220e3}$ (green). $t=0$ corresponds to the moment the pump power reaches $50\%$ of
        its mean maximum value.
    	The extracted delay $t_D$ of the orange signal is shown by the grey shaded area.
    	(b)~Measured mean delays of the first emitted SF burst vs. $N_\mu$ for $\Delta_p=18.4\Gamma$, with error bars representing the standard deviation.
    	The expected theoretical dependence [Eq.~\eqref{eqn:delay}] is shown
    	considering $N_\mu$ (dashed green line) as well as
    	the MCN $N_\text{mc}$ (solid orange line) for $\beta=0.07$.
	}
    \label{fig:example}
\end{figure}

In the following we study SF in the effective TLS shown in Fig.~\ref{fig:levelscheme}.
Before each measurement, we switch off the trapping potential to avoid
inhomogeneous broadening by the FORT.
When the pump is resonant with $|1\rangle \leftrightarrow |3\rangle$ we can
measure $N$ by time-resolved optical pumping \cite{PYH21}.
The measured shot-to-shot fluctuations of $N$ are $\lesssim \SI{3.5}{\percent}$. 
We then detune the pump field from
$|1\rangle \leftrightarrow |3\rangle$ by a variable amount $\Delta_p$.
As confirmed by measurements and a numerical simulation,
the polarization of the Stokes field is predominantly linear and orthogonal
to the pump field. We thus observe the temporally-resolved Stokes radiation by detecting light
orthogonally polarized to the pump using a polarizing beam splitter (PBS).  
We show in Fig.~\ref{fig:example}(a) exemplary data for the transmitted pump
power through the HCPBGF without atoms (blue) and for the transmitted Stokes
power with $N=\num{83e3}$ atoms loaded into the HCPBGF (orange) versus
time.
No Stokes light is detected for several hundred nanoseconds after the pump is switched on.
Then, a short burst of peak power $P_s$ is emitted,
followed by another burst of smaller peak power due to a coherent ringing \cite{GH82,HTF85}.
This trace, showing a large first burst followed sometimes by a smaller one,
represents the dominant temporal shape of the emitted Stokes light for
$N<\num{220e3}$ and agrees with earlier observations of SF \cite{GH82,HTF85,PBK08}. 
We note that the transmitted pump power shows complementary dynamics to the Stokes field, i.e., fast absorption, which was termed `superabsorption' in \cite{HBS14}.
For our maximum attainable $N=\num{220e3}$, however, the dominant shape changes [green trace in Fig.~\ref{fig:example}(a)].
Here, the ringing is more pronounced and the second burst is largest which is usually not the case \cite{RLW89}.
As to obtain good statistics for the Stokes signal, we repeat
the measurement for each set of experimental parameters $(N,\Delta_p)$
$100$ times (see video \cite{Peters2023}).
The Stokes signal exhibits huge shot-to-shot fluctuations in amplitude, delay, intensity and shape.
The burst dynamics occur on a timescale much shorter than expected for
the calculated single-atom scattering rate of
$\langle\Gamma_R\rangle_r=2\pi\times \qty{45}{\kHz}$, where
$\langle ...\rangle_r$ denotes a radial average over both the pump
intensity and atomic density distribution (see Appendix \ref{app:effective_scattering_rate}).
From each single dataset we extract the delay $t_D$ and peak power
$P_s$ of the first
emitted Stokes burst by an automated algorithm to obtain their mean values
$\langle t_D \rangle$ and $\langle P_s \rangle$ and the corresponding standard deviations.

Figure~\ref{fig:example}(b) shows exemplary data for the measured mean delays
vs. reduced atom number $N_\mu$ (symbols).
Here, $\Omega_p$ and $\Delta_p$ are the same as in Fig.~\ref{fig:example}(a) and we
expect the system to be well-described by an effective TLS. The standard deviations from the mean delay are a significant $\sim 38\%$, which is much larger than both the $\sim 4\%$ uncertainty of determining $t_D$ in a single shot and the expected uncertainty of $\sim 4\%$ due to fluctuating atom number and pump intensity (see Appendix \ref{app:parameters}). As the relative delay fluctuations from the mean $\langle t_D \rangle$ due to vacuum fluctuations can be estimated as $2.6/\ln{N}$ \cite{GH82} ($\sim 0.2..0.3$), we attribute the dominant contribution to the observed fluctuations to vacuum fluctuations initiating the decay process \cite{HKS79,RW84}.
We note that the standard error of the mean is a factor 10 smaller than the standard deviation shown by the error bars.
In the same plot we show the expected dependence (dashed green line) for a
homogeneous ensemble of Fresnel number $F \sim 1$ (see Appendix \ref{app:fresnel}) \cite{PSV79}
\begin{equation}
     \langle t_D(N) \rangle=\frac{1}{4\mu N \Gamma_R}\left[\ln{\sqrt{2\pi N}}\right]^2.
     \label{eqn:delay}
\end{equation}
The measured mean delays are larger than Eq.~\eqref{eqn:delay} predicts,
as expected for inhomogeneously-broadened systems \cite{HHK80}. 

\begin{figure}[bt]
    \centering
    \includegraphics[width=\columnwidth ]{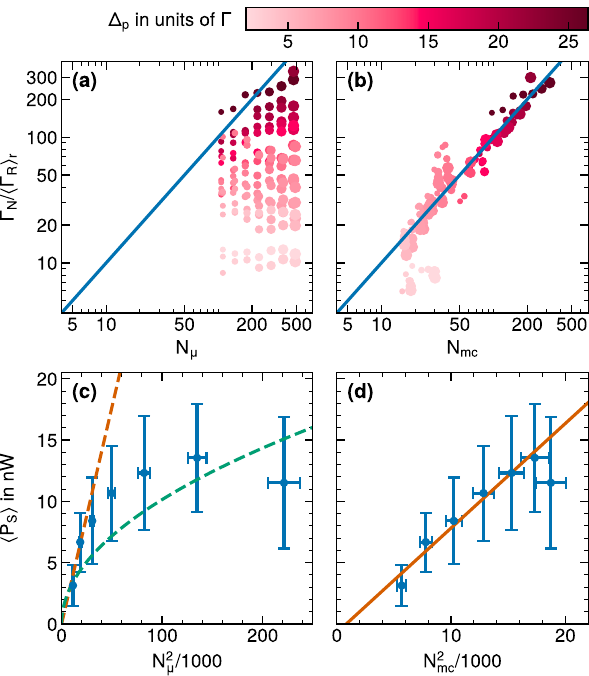}
    \caption{
    	(a,b)~Experimentally determined initial relative collective decay rate $\Gamma_N / \langle \Gamma_R \rangle_r$
    	vs. $N_\mu = \mu N$ (a) and MCN $N_\text{mc}= \eta \mu N$ (b) for
    	$2\Gamma \leq \Delta_p \leq 26.4\Gamma$ at $\Omega_p^{(0)}=6.4\Gamma$.
    	The symbol size is proportional to $N$. The solid blue lines show the dependence $\Gamma_N / \langle \Gamma_R \rangle_r = N_\mu$ (a) and $\Gamma_N / \langle \Gamma_R \rangle_r = N_\text{mc}$ (b) for homogeneous conditions. (c,d)~Measured mean peak power of the first SF burst vs. $N_\mu^2$ (c) and $N_\text{mc}^2$ (d) for $\Delta_p=18.4\Gamma$. The green (orange) dashed line in (c) represents a linear (quadratic) scaling with $N_\mu$ for reference.
    	The orange solid line in (d) represents a least-squares linear fit to the data, excluding the last data point (see text).
    }
    \label{fig:ratio}
\end{figure}

To extract the initial collective decay rate $\Gamma_N$ from our SF data, we cannot resort to determining the decay rate as, e.g., in linear-optics SR due to the non-exponential dynamics. An order-of-magnitude estimation can be done by measuring the temporal widths $\tau_b$ of the first SF bursts, as $\Gamma_N \propto \tau_b^{-1}$ is a reasonable assumption \cite{GH82}. However, as there is, to the best of our knowledge, no precise theoretical prediction for $\Gamma_N(\tau_b)$ in our considered case of SF bursts with ringing, we proceed as follows.
From Eq.~\protect\eqref{eqn:delay} we determine the collective decay rate as $\Gamma_N = [\ln{\sqrt{2\pi N}}]^2 / ( 4\langle t_D \rangle )$ using the measured $\langle t_D \rangle$ and $N$.
Fig.~\ref{fig:ratio}(a) depicts the measured relative collective decay rate $\Gamma_N/\langle\Gamma_R\rangle_r$ (symbols) vs. $N_\mu$ for a large range of detunings, along with the
expected theoretical dependence $\Gamma_N/\langle\Gamma_R\rangle_r = N_\mu$ (blue line), where  $\mu\approx \textrm{NA}^2/4=2.1\times 10^{-3}$.  
The collective decay rate is significantly enhanced by more than two orders of magnitude compared to the single-atom case. Only at the largest detunings it scales approximately linearly with $N_\mu$, but with a detuning-dependent slope.
Moreover, the experimental data does not collapse onto $\Gamma_N/\langle\Gamma_R\rangle_r=N_\mu$ expected for homogeneous conditions, except for the lowest $N$ and largest $\Delta_p$.
Close to resonance, $\Gamma_N/\langle\Gamma_R\rangle_r$ is approximately independent of $N$.
Thus, $N_\mu$ is not a good scaling parameter.
We note that we obtain quite similar values for $\Gamma_N/\langle\Gamma_R\rangle_r$ (differing by a factor of $\sim 1.7$) when determining $\Gamma_N$ from the burst widths $\tau_b$ instead of the mean delays $\langle t_D \rangle$ (see Appendix \ref{app:decay_rate_via_widths}), indicating that a determination of $\Gamma_N$ from $\langle t_D \rangle$ is indeed correct.
Next, we plot in Fig.~\ref{fig:ratio}(c) the measured mean peak Stokes power (symbols) as a function of $N_\mu^2$.
We observe a $N_\mu^2$ scaling (dashed orange line) only for small $N$.
As $N$ becomes larger, the scaling becomes approximately linear (dashed green line).
This change in scaling indicates a significant reduction in cooperativity for
increasing $N$. A similar behavior is seen for other detunings with a breakdown of $N_\mu^2$
scaling occurring for smaller $N$ as $\Delta_p$ becomes smaller.
Our results clearly demonstrate strong collective decay for our dilute, disordered
atomic ensemble coupled to an HCPBGF.
However, the in part significant deviations from theory for homogeneous conditions evoke a deeper investigation as to understand the limits to collective emission in our system and to optimize it.

\section{Determining the MCN}\label{sec:mcn}
Based on the concept of MCN \cite{AC70}, we determine in the following the fraction of
atoms $N_{\text{mc}} = \eta \mu N$ which can radiate collectively, i.e., can be viewed
as homogeneously driven.
We identify two major mechanisms affecting the MCN.
(i) Inhomogeneous spectral broadening, and (ii) pump attenuation.
Due to the radial as well as longitudinal dependence of the pump intensity determining decay rate, AC Stark shift, and Stokes gain, we factorize the problem: we first determine radially-averaged quantities and then apply them to average longitudinally (see the Appendix \ref{app:modeling_pump_attenuation} for further details).

First, we consider \textit{inhomogeneous spectral broadening}.
As noted in \cite{AC70}, the number of cooperative scatterers is reduced by
the ratio $\sigma_\text{hom}/\sigma_\text{inh}$ of (homogeneous) excitation
bandwidth $\sigma_\text{hom} \sim \tau_p^{-1}$
and inhomogeneous linewidth $\sigma_\text{inh}$
if $\sigma_\text{hom}<\sigma_\text{inh}$.
We estimate $\sigma_\text{inh} = \delta S + \delta\Gamma_R$, where $\delta S$ and $\delta \Gamma_R$ are standard deviations of the AC Stark shift $S$ and the effective scattering rate $\Gamma_R$ from their average values $\langle S \rangle_r$, $\langle \Gamma_R \rangle_r$. This yields a first correction factor 
\begin{equation}
    \eta_\text{inh}=\frac{1}{\tau_p\sigma_\text{inh}} 
\end{equation}
to the MCN.
Next, we consider \textit{pump attenuation} along the propagation direction
$z$ through the optically-dense ensemble.  
Such attenuation is detrimental to excitation of a timed-Dicke state \cite{BPC11,BBP13}, and in addition directly affects the decay rate of the effective TLS.
We first take (off-)resonant absorption according to the Beer-Lambert law into account by 
approximating the absorption coefficient by its initial value
\begin{equation}
    \alpha(\Delta_p)
    \approx \langle \alpha(\Delta_p,0) \rangle_r
    = \left\langle \alpha_0 \frac{\Gamma^2}{
        4\left[\Delta_p+ 2S(r,0)\right]^2} \right\rangle_r,
\end{equation}
where $\alpha_0$ is the resonant OD, and we performed a radial average. 
Noting further that the Stokes intensity eventually grows as $I_s(\Delta_p,z') \sim \exp[ G_s(\Delta_p)z' ]$ \cite{RM81,PSV79}, where $z'=z/L$ and the Stokes gain $G_s(\Delta_p,0) \approx \left\langle 2\alpha_0\Gamma_R(\Delta_p,0)/\gamma \right\rangle_r$, with $\gamma$ being the ground state decoherence rate, we next consider pump attenuation due to conversion into the Stokes field.
We thus make the ansatz
$I_p(\Delta_p,z') \approx I_p^{(0)}\exp[ -\tilde{\alpha}(\Delta_p)z' ]$ for the pump intensity, where the total attenuation factor is
\begin{equation}
    \tilde{\alpha}(\Delta_p)=\alpha(\Delta_p,0)+\beta(\Delta_p)G_s(\Delta_p,0).
\end{equation}
In a first approximation we assumed steady-state conditions for both absorption and conversion and that $\tilde{\alpha}(\Delta_p)$ is given by its initial radially-averaged value at $z'=0$. 
This assumption of steady-state conditions is a good approximation for pump absorption as the duration of transients is roughly given by $\Gamma^{-1} \ll t_D$. 
However, as the transient regime during buildup of the Stokes field is roughly determined by $\gamma^{-1} \sim t_D$, this ansatz is probably less accurate for pump conversion. 
We thus introduced a scaling factor $\beta(\Delta_p)$ for the Stokes gain as the only free parameter in our model (see Appendix \ref{app:beta} for a more detailed discussion). This parameter accounts for neglecting transient conditions during Stokes field buildup exhibiting a time-dependent Stokes intensity evolution \cite{RM81}.
We next calculate the effective collective decay rate as
\begin{equation}
    \left\langle \Gamma_N^\text{eff} \right\rangle_{r,z}
    = \mu N \left\langle \Gamma_R\left[\Omega_p(r, z')\right] \right\rangle_{r,z}
\end{equation}
by averaging along the $r-$ and $z-$coordinates and using $\Gamma_R\left[\Omega_p(r, z')\right]$ from Eq.~\eqref{eqn:scattering_rate} with
\begin{equation}
    \Omega_p(r,z') = \Omega_p^{(0)} e^{-\frac{r^2}{\sigma_p^2}} 
    e^{-\frac{1}{2}\tilde{\alpha}(\Delta_p)z'}.
\end{equation}
Following \cite{BPG16,KRH20} we can now determine a shadow factor
\begin{equation}
	\eta_S = \frac{\left\langle
		\Gamma_N^\text{eff} \right\rangle_{r,z}}{\mu N \langle\Gamma_R
	\rangle_r}
\end{equation}
which incorporates pump attenuation.
Here, $\langle\Gamma_R\rangle_r$ is the radially-averaged single-atom
scattering rate at $z'= 0$ used as a reference without attenuation.
Combining inhomogeneous broadening and pump attenuation, we finally obtain
the MCN as
\begin{equation}
    N_{\text{mc}}= \eta_\text{inh}\cdot \eta_s \cdot \mu \cdot N.
     \label{eqn:MCN} 
\end{equation}

\begin{figure}
    \centering
    \includegraphics[width=0.8\columnwidth ]{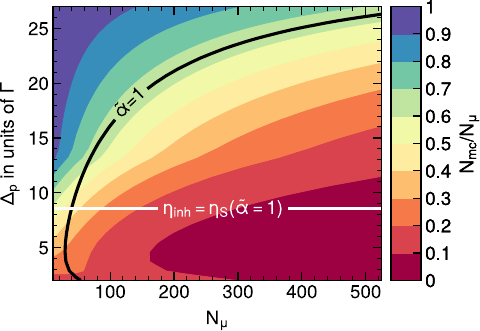}
    \caption{
    	Relative MCN $N_\text{mc}(\Delta_p,N)/N_\mu$ vs. $N_\mu$ and detuning $\Delta_p$.
    	The black line shows where $\tilde{\alpha}(\Delta_p,N)=1$ and the white line where $\eta_\text{inh}$ is as large as $\eta_s(\tilde{\alpha}{=}1)$.
	}
    \label{fig:relativeMCN}
\end{figure}

To test our simple model, we determine $\langle t_D(N_\text{mc})\rangle$ using $N_\text{mc}$
instead of $N_\mu$ in Eq.~\eqref{eqn:delay} by manually adjusting $\beta$
as the only free parameter to yield the best agreement with the experimental
data [see orange line in Fig.~\ref{fig:example}(b)].
In the range of detunings  $\Delta_p \gtrsim \Omega_p$ where our effective TLS
model should be valid, we determine $\beta(\Delta_p)=0.182-6.1\times10^{-3}\Delta_p/\Gamma$ from a linear fit.
Although we currently have no exact explanation for this dependence on $\Delta_p$, we suspect that it is due to a larger temporal ratio of transient to steady-state conditions at larger detunings (see Appendix \ref{app:beta} for more details).
For smaller $\Delta_p$, $\beta$ increases dramatically, indicating that our model is inapplicable in this regime.
We use $\beta(\Delta_p)$ to calculate $N_\text{mc}(\Delta_p,N)$ for all our data and plot the results for the relative MCN $N_\text{mc}(\Delta_p,N)/N_\mu$ in
Fig.~\ref{fig:relativeMCN}.
Only for the smallest $N$ and largest $\Delta_p$ the MCN is close to $N_\mu$.
As $N$ increases and $\Delta_p$ decreases, however, the MCN becomes smaller and only $\sim 10\%$ of the usual collective cooperativity $N_\mu$ contributes to collective scattering.
We can now plot $\Gamma_N(N_\text{mc})/\langle\Gamma_R\rangle_r$ and $\left\langle P_s(N_\text{mc}^2) \right\rangle$, respectively, to test their scaling [see Fig.~\ref{fig:ratio}(b,d)]. 
In contrast to Fig.~\ref{fig:ratio}(a) where $N_\mu$ was used, $\Gamma_N(N_\text{mc})/\langle\Gamma_R\rangle_r$ collapses onto the linear dependence $N_\text{mc}=\eta \mu N$ over a large range of detunings and atom numbers. Moreover, we also observe a quadratic scaling of $\langle P_s\rangle$ with the MCN over almost the full range of atom numbers.
Only at the largest $N_\text{mc}$ the scaling of $\langle P_s\rangle$ changes.
In this regime [green trace in Fig.~\ref{fig:example}(a)], however, the second instead of the first burst exhibits the largest power in the dominant fraction of measurements, which is not the case in standard SR theory \cite{GH82}.
Therefore, the breakdown of $N^2$ scaling is not surprising, and we excluded this data point from the fit.

\section{Discussion}\label{sec:discussion}
Our results demonstrate that the concept of MCN can recover the scaling of collective decay in homogeneous ensembles for inhomogeneous conditions. This provides insight into the limits of collective decay in our setup.
To this end we plot in Fig.~\ref{fig:relativeMCN} the boundaries for which pump attenuation (black line) \cite{BBP13,AKK16}
and inhomogeneous broadening (white line) are becoming detrimental to collective decay. 
Our system is obviously limited by pump attenuation for large detunings.
Only for $\Delta_p\lesssim 9\Gamma$ inhomogeneous spectral broadening is becoming relevant.
Besides supporting the picture of linear-optics SR presented in \cite{WCK21,ACG22}, our results demonstrate that one can find a regime that renders the radial inhomogeneities insignificant for atomic ensembles in a hollow-core fiber.
For appropriately large detunings, almost all atoms contribute to collective scattering, leading to a pronounced enhancement of the collective decay rate, at least for negligible dispersion.

Despite its simplicity, our model seems to capture the relevant physics.
Only for detunings $\Delta_p \lesssim 7\Gamma$, where the approximation of an effective TLS is not valid anymore for $\Omega_p^{(0)} = 6.4\Gamma$, deviations from the experimental data become more significant.
Furthermore, an estimation shows that dispersion is likely to become relevant here as
frequency components of the SF bursts are lying within the dispersive bandwidth of the excited state \cite{WCK21}.
This is currently not captured by our model.

\section{Conclusion and Outlook}\label{sec:summary}
We presented an experimental study of superfluorescence in a disordered ensemble of cold atoms coupled to a hollow-core fiber. We demonstrated a decay rate collectively-enhanced by more than two orders of magnitude for $N\lesssim\num{2.2e5}$ atoms. By implementing an effective TLS we were able to study the decay with negligible dispersion for a large range of detunings and optical depths. Thereby we could tune the amount of inhomogeneity, studying its effect on the SF decay and developing a simple theoretical model considering inhomogeneous broadening and attenuation of the excitation field. Using this model, we determined a maximum cooperation number $N_\text{mc}$ of atoms that can be treated homogeneously and thus decay collectively. 
We demonstrated that using $N_\text{mc}$ instead of $\mu N$, the $N$ scaling known to collective decay in homogeneous ensembles could be recovered. This allowed for a physical understanding of the limits to enhanced collective decay in our waveguide-coupled ensemble. 
Our results provide a simple tool to understand and optimize collective scattering in inhomogeneous extended ensembles, as can be found, e.g., in astrophysics \cite{HMR18}, solids \cite{CZW16,BJB17}, waveguide QED \cite{SPI23}, novel atomic clocks \cite{MYC09,NWC16}, and quantum optics \cite{AMA17}.

Future studies could try determining a further scaling factor for the MCN incorporating dispersion which could help finding an analytical model of collective decay for dispersive ensembles \cite{GK17}.
Also, applying the here demonstrated approach to other types of experimental systems exhibiting inhomogeneities would be interesting as to, e.g., determine their limits to support collective decay.
Furthermore, it might be interesting to study the regime where the SF bursts exhibit an unusual dominant temporal shape  [see Fig.~\ref{fig:example}(a)].
Finally, our demonstration of collectively-enhanced Stokes scattering rates could be used during the pair-generation process in quantum light sources based on spontaneous four-wave mixing \cite{BHS22} thereby reducing their required pump power.

\begin{acknowledgments}
The authors would like to thank W.~Guerin and L.P.~Yatsenko for discussions, E.~Giese, R.~Walser, and V.~Stojanovic 	for discussions and comments on the manuscript, and T.~Halfmann for his	continuous support and comments on the manuscript.
We acknowledge the group of T.~Walther for providing us with their 	home-built ultra-low noise laser diode driver with high modulation bandwidth.
The project received funding from the European Union’s Horizon 2020 research and innovation programme under the Marie Skłodowska-Curie grant 	agreement No.~765075 and by the Deutsche Forschungsgemeinschaft (DFG, German Research Foundation) project number 410249930.
\end{acknowledgments}

\appendix

\section{Determination of Experimental Parameters for the Theoretical Model}
\label{app:parameters}

Comparing the experimental data to the theoretical model requires knowledge of several experimental parameters. 
The geometric factor describing the fraction of spontaneous light emitted into the fiber core mode is approximately given by $\mu \approx \textrm{NA}^2/4$, where $\textrm{NA}$ is the numerical aperture of the HCPBGF. Taking the measured $1/e^2$ mode field radius $\sigma_p=\qty{2.75(20)}{\um}$ of the near-Gaussian-shaped intensity distribution, we determine $\textrm{NA}=\lambda/(\pi\sigma_p) = 0.092(6)$ with $\lambda=\qty{795}{\nm}$. This results in $\mu=2.1(3)\times 10^{-3}$.
The decay rate of the effective TLS depends on the branching ratio $R_B=\num{0.5}$ \cite{Steck2019} into the ground state $|2\rangle$, the Rabi frequency $\Omega_p$ of the pump, and its detuning $\Delta_p$, as does the AC Stark shift $S$. We measure the pump power ($\sim$\SI{5}{\percent} shot-to-shot fluctuations) at a certain location inside the experimental setup and correct for transmission losses from the HCPBGF to this location. Then, using the measured mode field diameter, we calculate the peak Rabi frequency $\Omega_p^{(0)}$ according to the parameters given in \cite{Steck2019}. The detuning $\Delta_p$ is controlled by the frequency of the acousto-optic modulator (AOM) modulating the pump with a rise time $\tau_p=\qty{130(5)}{\ns}$. The frequency uncertainty of the AOM is negligible leaving only the uncertainty of the pump laser lock-point. We estimate this uncertainty as $\lesssim \Gamma/10$, which can therefore be neglected as well in our analysis.
The total atom number $N$ is determined as described in Sec.~\ref{sec:expstudies} with a shot-to-shot uncertainty of \SI{3.5}{\percent}. The radial atomic density profile was investigated in \cite{PYH21} from which we assume a Gaussian-shaped radial density with a $1/e$ half width of $\sigma_a=\qty{1.7}{\um}$. Using an atomic density $\mathcal{N}(r,z)=\mathcal{N}(r)$ and $\Omega_p(r)$ we determine the peak optical depth as
\begin{equation}
     \alpha_0 = \frac{4}{\sigma_p^2} L \int_0^{r_c} dr \, r \, \sigma_{13} \, \mathcal{N}(r) \, e^{-\frac{2r^2}{\sigma_p^2}},
\end{equation}
where $\sigma_{13}$ is the resonant absorption cross-section of the transition $|1\rangle \leftrightarrow |3\rangle$ according to \cite{Steck2019}, $r_c$ is the core radius of the HCPBGF and we determined $\mathcal{N}(r)$ such that $2\pi \, L \int dr \, r \, \mathcal{N}(r) = N$.
As the final parameter to determine, there is a residual decoherence rate between the two ground states due to residual magnetic field gradients and time-of-flight broadening once the guiding potential is off for the measurements. This value was determined from light storage and retrieval experiments as $\gamma_0\sim 0.057(6)\Gamma$ \cite{BHS22}.

\section{Alternative Method of Determining the Collective Decay Rate}
\label{app:decay_rate_via_widths}

\begin{figure}[t]
    \centering
    \includegraphics[width=8cm]{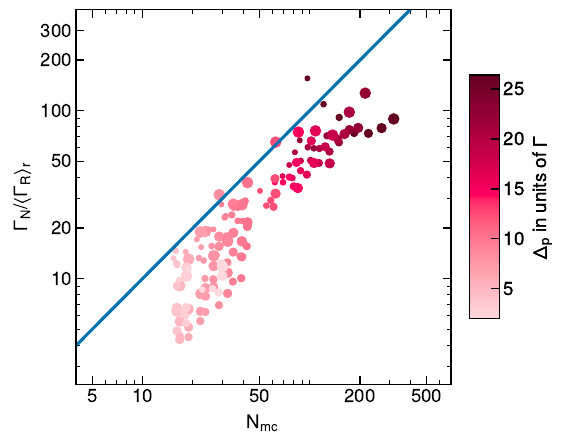}
    \caption{Same as Fig.~\ref{fig:ratio}(b) but using the widths of the first bursts instead of the mean delays to determine $\Gamma_N$.}
    \label{fig:ratio_widths}
\end{figure}

Determination of the collective decay rate for our experimental data is not as straightforward as, e.g., in the linear-optics SR regime, where a fit to the exponentially decreasing detected power yields the decay rate \cite{AKK16}. In our case of SF with strong bursts and ringing we thus have to apply another method. As discussed in Sec.~\ref{sec:expstudies}, we determined $\Gamma_N$ from the mean delays $\langle t_D \rangle$ of the first SF bursts for which a theoretical prediction exists [see Eq.~\eqref{eqn:delay}].
As this method is somewhat indirect, we apply in the following an alternative method.
As was done, e.g., by Okaba \textit{et al.} in the case of SR with bursts and ringing \cite{OYV19}, the width of the first burst can be used as a measure for the initial collective decay rate.
In the case of small-sample SR with a single burst the relation between the temporal burst width $\tau_b$ (FWHM) and the initial collective decay rate is given by $\Gamma_N=3.5/\tau_b$ \cite{GH82}. For large-sample SR/SF, corresponding to our experiment, we are, however, not aware of an analogous derivation for $\Gamma_N(\tau_b)$.
Despite the lack of of a precise theoretical prediction, we determined $\Gamma_N$ from our data using the first burst widths and the small-sample relation $\Gamma_N=3.5/\tau_b$.
The results for $\Gamma_N/\langle\Gamma_R\rangle_r$ vs. $N_\text{mc}$ are plotted in Fig.~\ref{fig:ratio_widths}.
Comparing this plot to Fig.~\ref{fig:ratio}(b), we note the following: (a) The order of magnitude of $\Gamma_N$ is the same. (b) Plotting $\Gamma_N$ vs. the MCN recovers a near-linear scaling in both cases. (c) Scaling $\Gamma_N$ in Fig.~\ref{fig:ratio_widths} by a factor of $\sim 1.7$ results in a scattering of the datapoints around $\Gamma_N/\langle\Gamma_R\rangle_r = N_\text{mc}$ (blue line). (d) The scattering of the data is larger when using the burst widths compared to the mean delays.

In view of these results we conclude that the initial collective decay rate of SF can be determined either from the mean burst delays or from their temporal widths. The former method produces more reliable results and provides a direct comparison to theory. Using rather the burst delays instead of their widths is also suggested by the uncertainties in the fitting process. Our data shows that the delays can be determined with a smaller relative uncertainty than the widths. This is intuitive as the width is also dependent on the fitted amplitude. 
Finally we note that the widths determined from the fits to the single-shot data are in general larger than the raw data would suggest. This might explain why the datapoints in Fig.~\ref{fig:ratio_widths} are below the theoretical line. Another reason for this discrepancy could, however, be that the relation $\Gamma_N=3.5/\tau_b$, valid for small samples, cannot be directly applied to the large-sample case.

\section{Determination of the Maximum Cooperation Number}
\label{app:mcn}

In the following we present the model used to calculate the maximum cooperation number $N_\textrm{mc}$. 
It is mainly based on the ideas presented by Arecchi and Courtens \cite{AC70}, Bienaimé \textit{et al.} \cite{BBP13}, and Bachelard \textit{et al.} \cite{BPG16}. 
First, inhomogeneous broadening reduces the number of cooperative atoms \cite{AC70}.
Second, attenuation of the pump field as it propagates through the optically-dense ensemble reduces the effective scattering/decay rate $\Gamma_N$, which can be viewed as a reduction of the number of synchronized scatterers.
We account for this by calculating a factor $\eta_s$ from an effective scattering rate in the spirit of the shadow effect introduced by Bachelard \textit{et al.} \cite{BPG16}. We note that in our case the decay rate $\Gamma_R$ itself depends on the (attenuated) excitation field, which is in contrast to decay in a true TLS where the decay rate is constant and only the excitation is inhomogeneous.

To find a simple analytical model, we make several assumptions, as the actual system exhibits complicated nonlinear spatial and temporal dynamics. First, there is a radial variation of the atomic density distribution $\mathcal{N}(r)=\mathcal{N}_0\exp \left(-r^2/\sigma_a^2 \right)$, and we assume a constant density along the propagation direction $z$. Second, the pump intensity $I_p(r,z) \propto \Omega_p^2(r,z)$, determining both Raman scattering rate and AC Stark shift of the ground state, is potentially attenuated along the propagation direction $z$ and also exhibits a radial variation $\Omega_p^2(r,0) = \Omega_p^2(0,0)\exp \left( -2r^2/\sigma_p^2 \right)$ . Its $1/e^2$ half width is only slightly larger than $\sigma_a$.
We use the fact that $\Omega_p^2(r,z)$ can be factorized as 
\begin{equation}
 \Omega_p^2(r,z) = \Omega_p^2(0,0) \cdot g_p(r) \cdot h_p(z),
\end{equation} 
Furthermore, inhomogeneous broadening due to the radial variations is expected to dominate over longitudinal variations as it is also present for large detunings.
These assumptions allow for considering the radial and longitudinal variations independently. In the following we start by calculating the radial averages weighed by the atomic density distribution at the start of the medium at $z=0$ for obtaining the inhomogeneous linewidth. Then, we take the radial and longitudinal average to determine the effective scattering rate. Using these results we then determine a MCN.

\subsection{Inhomogeneous Broadening}
\label{app:inhbroad}

As discussed by Arecchi and Courtens, inhomogeneous broadening reduces the number of cooperative atoms by the ratio of (homogeneous) excitation bandwidth $\sigma_\text{hom}$ to inhomogeneous linewidth $\sigma_\text{inh}$ \cite{AC70}. The excitation bandwidth can be estimated by the rise time $\tau_p$ of the pump field, $\sigma_\text{hom}=1/\tau_p$. 
The inhomogeneous linewidth of the effective TLS can be estimated as follows.
The pump intensity inside the HCPBGF determines the AC Stark shift $S$ of the ground state $F{=}1$
\begin{equation}
    S(r,z) = \frac{\Omega_p^2(r,z)}{4\Delta_p}
    \label{eqn:ACS}
\end{equation}
and the decay of the effective TLS
\begin{equation}
    \Gamma_R(r,z) = \Gamma\frac{\Omega_p^2(r,z)}{8\left[\Delta_p+2S(r,z)\right]^2}.
\end{equation}

We therefore calculate the radial averages of $\Omega_p^2(r,0)$, $S(r,0)$, and $\Gamma_R(r,0)$ at the beginning $z=0$ of the medium, weighed by the atomic density distribution as 
\begin{equation}
    \langle f(r) \rangle_r = \frac{2}{\sigma_a^2} \int_0^\infty dr\, r \, e^{-\frac{r^2}{\sigma_a^2}} f(r),
\end{equation}
where $f(r)=\left[\Omega_p^2(r,0),\, S(r,0),\, \Gamma_R(r,0)\right]$.

Using the radially-averaged values, we now estimate the inhomogeneous linewidth $\sigma_\text{inh}$ as follows. 
Due to the radial Gaussian dependence of the strong pump field, atoms in the ensemble of radial width $\sigma_a$ will experience an inhomogeneous AC Stark shift $S(r)$ of the ground state $F{=}1$  as well as an inhomogeneous decay rate $\Gamma_R(r)$. 
As a measure for the corresponding linewidth we calculate the standard deviation
\begin{equation}
    \delta f = \sqrt{\left\langle \left[ f(r) - \langle f(r) \rangle_r \right]^2\right\rangle_r}\label{eqn:stdev}
\end{equation}
for $f(r)= \left[S(r,0),\, \Gamma_R(r,0) \right]$ and set
\begin{equation}
    \sigma_\text{inh} = \delta S + \delta\Gamma_R.
\end{equation}

\begin{figure*}[t]
    \centering
    \includegraphics[width=\textwidth]{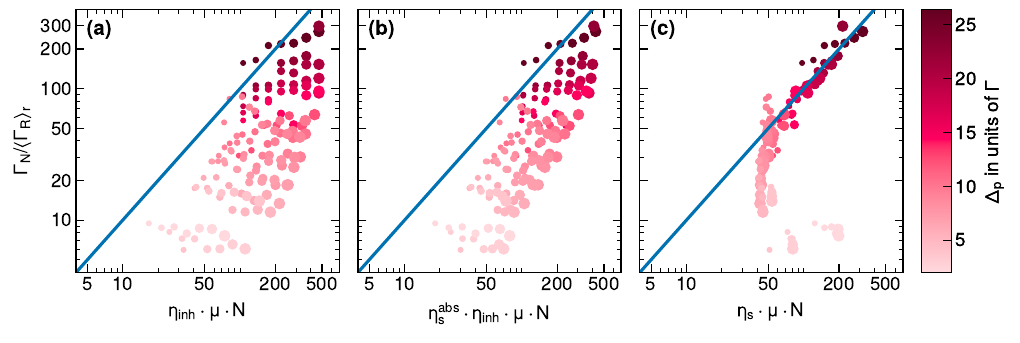}
    \caption{Initial relative collective decay rate $\Gamma_N / \langle\Gamma_R\rangle_r$ vs. corresponding MCN when considering only inhomogeneous broadening (a), inhomogeneous broadening and pump absorption (without conversion) (b), pump attenuation (absorption and conversion) without inhomogeneous broadening (c). Compare to Fig.~\ref{fig:ratio}(a,b).}
    \label{fig:ratios}
\end{figure*}

Using the measured rise time $\tau_p$ of the pump pulse, we obtain the first of two scaling factors, \begin{equation}
    \eta_\text{inh}=\frac{1}{\tau_p\sigma_\text{inh}},
\end{equation}
to calculate the MCN. 
Note that we only apply this scaling factor if the inhomogeneous bandwidth is larger than the homogeneous bandwidth.
Considering only inhomogeneous broadening in the MCN, we obtain the results shown in Fig.~\ref{fig:ratios}(a) which have to be compared to those shown in Fig.~\ref{fig:ratio}(a).
Obviously, taking inhomogeneous broadening into account slightly improves the scaling but is only relevant in the range of $\Delta_p\lesssim 9\Gamma$.

\subsection{Effective Scattering Rate}
\label{app:effective_scattering_rate}

Having determined a scaling factor which accounts for inhomogeneous broadening (in the radial direction), we next turn to the variations in the longitudinal direction $z$. This is relevant for our considered system, as propagation through the optically-dense atomic ensemble can attenuate the pump field. To this end, we calculate an effective collective scattering rate weighed by the radial atomic density distribution $\mathcal{N}(r)$
\begin{equation}
\begin{aligned}
    \left\langle \Gamma_N^\text{eff} \right\rangle_{r,z}
    &= \mu N \Gamma \cdot \frac{2}{\sigma_a^2} \int_0^1 dz'  \int_0^\infty  dr \, r \, e^{-\frac{r^2}{\sigma_a^2}} \times \\
    &\quad \times \frac{\Omega_p^2(r,z')}{8\left[\Delta_p+2S(r,z')\right]^2},\label{eqn:effscattrate2}
    \end{aligned}
\end{equation}
where
\begin{equation}
    \Omega_p(r,z') = \Omega_p^{(0)} e^{-\frac{r^2}{\sigma_p^2}} e^{-\frac{1}{2}\tilde{\alpha}(\Delta_p)z'},
\end{equation}
and $\Omega_p^{(0)}$ $[=\Omega_p(0,0)]$ is the peak Rabi frequency of the pump. We assume attenuation of the pump intensity according to Beer-Lambert's law along the dimensionless distance $z'=z/L$ with an attenuation factor $\tilde{\alpha}(\Delta_p)$ affecting also $S(r,z')$ in Eq.~\eqref{eqn:ACS}. 
We obtain the corresponding collective scattering rate $\langle\Gamma_N\rangle_r$ without pump attenuation by setting $\tilde{\alpha}(\Delta_p)\equiv 0$ in Eq.~\eqref{eqn:effscattrate2}.
Similar to Bachelard \textit{et al.}, where a \textit{shadow factor} was defined to determine an effective reduction in the total scattering cross-section of an optically-dense ensemble \cite{BPG16}, we now define a shadow factor
\begin{equation}
    \eta_s=\frac{\left\langle \Gamma_N^\text{eff} \right\rangle_{r,z}}{\mu N\langle\Gamma_R\rangle_r},
\end{equation}
where $\langle\Gamma_R\rangle_r$ is the radially-averaged single-atom scattering rate at $z'=0$ used as reference without attenuation. $\eta_s$ is thus a measure for the reduction of the collective scattering rate due to pump attenuation. As $\left\langle \Gamma_N^\text{eff} \right\rangle_{r,z}\propto N$ we identify $\eta_s \mu N$ as an effective number of atoms that will scatter collectively when considering attenuation of the pump field.

\subsection{Maximum Cooperation Number}

Finally, combining inhomogeneous broadening, longitudinal pump attenuation, and the geometric factor $\mu$ we determine the MCN as
\begin{equation}
    N_\textrm{mc} = \eta_\text{inh}\cdot\eta_s \cdot \mu \cdot N.
\end{equation}

\section{Modeling Pump Attenuation}
\label{app:modeling_pump_attenuation}

In order to apply the general model described above to our experimental data, we need to identify which factors cause attenuation of the pump, i.e., we need to model the (dimensionless) attenuation factor $\tilde{\alpha}(\Delta_p)$. The first obvious consideration is (off-)resonant absorption. Our experiment allows for loading $N\lesssim2.2\times 10^5$ atoms into the HCF. With an OD per atom of $\alpha^* = 2.75\times 10^{-3}$ on the transition $^2S_{1/2}\, F{=}1 \leftrightarrow\, ^2P_{1/2}\, F'{=}2$ we therefore achieve resonant optical depths of $\alpha_0 \lesssim 600$. The detuning-dependent optical depth at $z = 0$ for large detunings is approximately given by
\begin{equation}
    \alpha(\Delta_p) \approx \alpha_0 \left\langle
        \frac{\Gamma^2}{4\left[\Delta_p+2S(r,0)\right]^2}
    \right\rangle_r,
\end{equation}
where we have taken a radial average (see Appendix \ref{app:inhbroad}).
For our experimental parameters we reach $\alpha(\Delta_p) \gtrsim 1$ for $\Delta_p \lesssim 12\Gamma$.  Setting $\tilde{\alpha}(\Delta_p) \equiv \alpha(\Delta_p)$ in Eq.~\eqref{eqn:effscattrate2}, where we assumed in a first approximation that the attenuation factor does not depend on $z$ and is given by its initial value at $z=0$,
yields a MCN $N_\textrm{mc}^\text{abs}=\eta_\text{inh}\cdot\eta_s^\text{abs} \cdot \mu \cdot N$ for each data point in Fig.~\ref{fig:ratio}(a). 
Plotting the initial relative collective decay rate $\Gamma_N / \langle\Gamma_R\rangle_r$ vs. $N_\textrm{mc}^\text{abs}$ we obtain the plot shown in Fig.~\ref{fig:ratios}(b).
Note that so far there is no free fitting parameter. Compared to Fig.~\ref{fig:ratios}(a), where only inhomogeneous broadening was considered, $N_\textrm{mc}^\text{abs}$ is a further improved scaling parameter compared to the collective cooperativity $N_\mu$. Yet the data points do not collapse onto the blue line showing $\Gamma_N / \langle\Gamma_R\rangle_r = N_\textrm{mc}^\text{abs}$. 

We therefore consider additional attenuation of the pump.
To this end we note that the Stokes buildup happens on a timescale determined by the Stokes gain \cite{RM81} and that the derivation of Eq.~\eqref{eqn:delay} was done in a regime of exponential growth of the SF burst \cite{PSV79}. 
We thus make the ansatz 
\begin{equation}
    I_p(z')\approx I_p^{(0)} e^{-{\tilde{\alpha}(\Delta_p)}z'},
\end{equation} 
where the total attenuation factor is given by
\begin{equation}
    \tilde{\alpha}(\Delta_p) = \alpha(\Delta_p) + \beta(\Delta_p)\cdot G_s(\Delta_p,0).
\end{equation}
We thus assume that the pump intensity $I_p$ decreases exponentially due to absorption \textit{and} conversion into the exponentially-growing Stokes field $I_s(z')\propto \exp{[\beta(\Delta_p)G_s(\Delta_p,0)z']}$ with a scaling factor $\beta(\Delta_p)$.

The initial radially-averaged Stokes gain can be approximately written as
\begin{equation}
    G_s(\Delta_p,z'{=}0) \approx \alpha_0 \left\langle
        \frac{2\Gamma_R(\Delta_p,0)}{\gamma}
    \right\rangle_r,
    \label{eqn:StokesGain2}
\end{equation}
where $\gamma$ is the ground state decoherence rate. For our all our experimental parameters we determine $G_s(\Delta_p,z'{=}0)>1$. This ansatz corresponds to steady-state conditions in \cite{RM81} while neglecting transient conditions during Stokes field buildup, as during the latter one the Stokes intensity evolution is additionally time-dependent which complicates the theoretical description \cite{RM81}. 
To account for this simplification we introduced the scaling factor $\beta(\Delta_p)$ for the Stokes gain, which is allowed to vary with $\Delta_p$ (see Appendix \ref{app:beta}).
We note that this scaling factor accounting for partially transient conditions during the buildup of coherence is only applied to the Stokes gain but not for absorption. This is motivated as follows: Although absorption also exhibits a transient after switch-on of the pump, the timescale is determined by the excited state decay rate $\Gamma$, corresponding to $\sim 30~$ns which is much smaller than the timescale of SF emission. On the other hand, the transient timescale for generation of the Stokes field is determined by the ground state coherence time $\gamma^{-1}$ \cite{RM81}, which is comparable to the timescale of SF emission. Thus, transient conditions are likely more relevant in the gain than in the absorption term.

In order to calculate the initial Stokes gain, we have to determine the decoherence rate $\gamma$. We identify three major contributions to $\gamma$ for our effective TLS. First, there is a residual decoherence rate between the two ground states determined by residual magnetic field gradients and time-of-flight broadening. This value was determined from light storage and retrieval experiments as $\gamma_0\sim 0.057(6)\Gamma$ \cite{BHS22}. Second, the inhomogeneous AC Stark shift of the the ground state by the pump field causes decoherence. We approximate this rate by the previously calculated standard deviation $\delta S$ (see Appendix \ref{app:inhbroad}). Third, the Raman scattering rate $\Gamma_R$ leads to decoherence due to decay between the two ground states. Therefore, we set the total decoherence rate as
\begin{equation}
    \gamma = \gamma_0 + \delta S + \langle \Gamma_R\rangle_r.
\end{equation}

Combining these results, we determine the MCN $N_\textrm{mc}$ for all measurement parameters $(\Delta_p,N)$. The results are shown in Fig.~\ref{fig:relativeMCN}.
Complementary to Fig.~\ref{fig:ratios}(a), where we only considered inhomogeneous broadening, we plot in Fig.~\ref{fig:ratios}(c) the results when considering only pump attenuation while neglecting inhomogeneous broadening. As we can see, attenuation of the pump field is relevant across the whole range of detunings, however, it fails to provide the correct scaling for $\Delta_p\lesssim 9\Gamma$. Only a combination of pump attenuation and inhomogeneous broadening yields the expected scaling.

\section{Scaling Factor $\beta$}
\label{app:beta}

\begin{figure}[t]
    \centering
    \includegraphics[width=8cm]{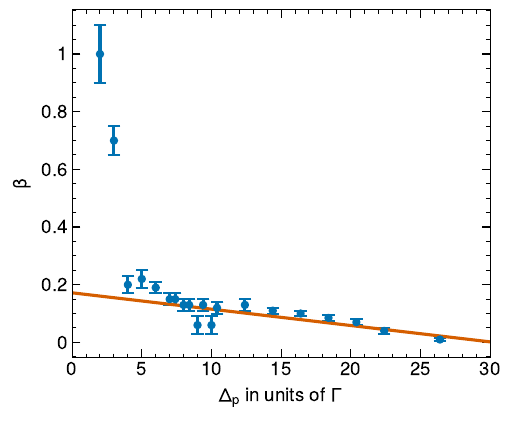}
    \caption{Scaling factor $\beta(\Delta_p)$ as determined from manual fits to the measured $\langle t_D(N) \rangle$. The solid line shows a linear fit for $\Delta_p > 6\Gamma$ with $\beta(\Delta_p) = 0.182 - 6.1 \times 10^{-3} \Delta_p/\Gamma$.}
    \label{fig:beta}
\end{figure}

The scaling factor $\beta$ is the only free parameter in our model introduced to account for the transient regime during buildup of the Stokes field. In this regime, the Stokes intensity does not simply grow exponentially as $\exp{(G_sz')}$ as in the steady-state regime, but exhibits a time dependence \cite{RM81} which is difficult to handle in our simple model. Using a constant $\beta \sim 0.07$ for all detunings and atom numbers works already satisfactory for detunings $\Delta_p > 7\Gamma$. However, not all measured delays are as well reproduced as shown in Fig.~\ref{fig:example}(b) for $\Delta_p = 18.4\Gamma$. As $\beta$ accounts for temporal dynamics during Stokes field buildup, it seems reasonable to assume that $\beta$ varies with the detuning (and atom number). 
Therefore, we plotted $\langle t_D(N) \rangle$ and determined $\beta(\Delta_p)$ by manually fitting the calculated mean delays [Eq.~\eqref{eqn:delay}] to the data. The resultant $\beta(\Delta_p)$ (see Fig.~\ref{fig:beta}) varies approximately linearly for $\Delta_p > 6\Gamma$.
The exact reason for such a linear dependence is currently unknown and we plan to further investigate it in the future trying to find a model for $\beta$. We currently speculate that the variation is due to a change in the relative time spent in the transient regime: Our experimental data shows that the mean delay $\langle t_D \rangle$ increases continuously from $\sim 0.7\gamma_0^{-1}$ at $\Delta_p=6\Gamma$ to $\sim 0.95\gamma_0^{-1}$ at $\Delta_p=26.4\Gamma$. Thus, it seems plausible that at larger detunings the temporal ratio of transient to steady-state conditions during coherence buildup gets larger, resulting in a smaller $\beta$.
For $\Delta_p \lesssim 6\Gamma$, $\beta$ increases strongly up to 1, indicating that our model is not valid anymore in this range. 
This comes as no surprise as this is approximately the regime where the effective TLS model breaks down, i.e., $\Delta_p \lesssim \Omega_p$ and dispersion should become relevant.
For calculating the MCN (see Sec.~\ref{sec:mcn}), we fit $\beta(\Delta_p)$ for $\Delta_p > 6\Gamma$ with a linear function yielding $\beta(\Delta_p) = 0.182 - 6.1 \times 10^{-3} \Delta_p/\Gamma$. We use this $\beta(\Delta_p)$ dependence for all measured detunings.


\section{Validity of the Pencil-Shaped Model for an Ensemble Coupled to a HCPBGF}
\label{app:fresnel}

The pencil-shaped model can be applied when the field propagates along a single direction and transverse field variations can be neglected \cite{GH82}. In free space, this is the case when the Fresnel number $F = \pi\sigma_a^2/(\lambda L) \sim 1$. 
In our case, the Fresnel number is $F \sim 10^{-4} \ll 1$ so a 3D description would be appropriate. However, the waveguide collects radiation emitted into a cone of opening angle $\textrm{NA} = \lambda/(\pi\sigma_p)$. Thus, our effective Fresnel number is $F'\sim \sigma_a/\sigma_p \sim 1$ and the pencil-shaped model can be applied. 


%

\end{document}